\documentclass[twocolumn,english,amssymb,aps,superscriptaddress,showpacs,amsmath,showkeys,floatfix,pra]{revtex4-1}
\usepackage[T1]{fontenc}
\usepackage[latin9]{inputenc}
\usepackage{geometry}
\usepackage[active]{srcltx}
\usepackage{amsmath}
\usepackage{graphicx}
\usepackage{esint}
\usepackage{epsfig}
\usepackage{epstopdf}
\usepackage{physics}
\usepackage{color}
\usepackage{natbib}
\usepackage{upgreek}
\usepackage{mathtools}
\usepackage{soul}
\usepackage{lipsum}
\usepackage{enumitem}
\usepackage{amsfonts}
\usepackage{eurosym}
\usepackage{braket}
\usepackage{setspace}
\newcommand {\rttensor}[1]{\overline{\overline{#1}}}
\newcommand\identity{1\kern-0.25em\text{l}}
\newcommand{\bs}{\boldsymbol}
\usepackage[dvipsnames]{xcolor}
\usepackage{ulem}
\usepackage{color}
\definecolor{blue}{rgb}{0,0,1}
\definecolor{green}{rgb}{0,1,0}
\definecolor{purple}{rgb}{0.5,0,1}
\makeatletter
\usepackage{babel}

\makeatother

\usepackage{babel}
\begin{document}


\title{Long-range interactions revealed by collective spin noise spectra in atomic vapors}

\author{J. Delpy}
\affiliation{Universit\'e Paris-Saclay, CNRS, Ecole Normale Sup\'erieure Paris-Saclay, CentraleSup\'elec, LuMIn, Orsay, France}
\author{N. Fayard}
\affiliation{Universit\'e Paris-Saclay, CNRS, Ecole Normale Sup\'erieure Paris-Saclay, CentraleSup\'elec, LuMIn, Orsay, France}
\author{F. Bretenaker}
\affiliation{Universit\'e Paris-Saclay, CNRS, Ecole Normale Sup\'erieure Paris-Saclay, CentraleSup\'elec, LuMIn, Orsay, France}
\author{F. Goldfarb}
\affiliation{Universit\'e Paris-Saclay, CNRS, Ecole Normale Sup\'erieure Paris-Saclay, CentraleSup\'elec, LuMIn, Orsay, France}

\begin{abstract}

{We report anomalous features in the spin noise spectroscopy (SNS) of a thin cell of a dense vapor of alkali atoms. At high densities and close to resonance, we observe a dramatic broadening of the spin noise spectra as well as an unexpected extra low-frequency noise component. With the help of a two-body model and simulations, we show that these features are the hallmark of a strong, long-range dipole-dipole interaction within the ensemble. The additional low-frequency noise reveals the correlated evolution of pair of atoms beyond the impact approximation. In this regime, we demonstrate that spin noise can no longer be obtained from one-body dynamics, opening the way for the characterization of many-body spin noise, atomic entanglement or higher order spin correlators in atomic vapors using SNS.}


\end{abstract}

\maketitle

Atomic vapors have become mainstream platforms for precision sensing and quantum technologies such as magnetometry \cite{budker_optical_2007}, quantum storage \cite{katz_strong_2022}, measurement of fundamental constants and testbeds for new physical theories \cite{smiciklas_new_2011, almasi_new_2020}. In such particle ensembles, interactions are ubiquitous and rule the dynamics of a wide variety of coherent and incoherent effects. For instance, in alkali vapors, spin-exchange (SE) collisions play a fundamental role in the dynamics of hyperfine and Zeeman coherences \cite{happer_effect_1977} and thus impact various physical systems including optically pumped magnetometers (OPM) \cite{allred_high-sensitivity_2002, savukov_effects_2005} or medical imaging techniques \cite{albert_biological_1994, navon_enhancement_1996}.

On the other hand, the optical measurement of spontaneous fluctuations of angular momentum in particle ensembles, called spin noise spectroscopy (SNS), has become a well known approach for non-invasive study of ground-state energy structures and dynamics \cite{aleksandrov_magnetic_1981, crooker_spectroscopy_2004, mihaila_quantitative_2006, zapasskii_optical_2013, swar_measurements_2018, swar_detection_2021}. 
Consequently, SNS was used during the last decade for extensive characterizations of SE collisions in single-species systems \cite{katsoprinakis_measurement_2007, mouloudakis_quantum_2019, mouloudakis_effects_2022} and two-species systems \cite{mouloudakis_interspecies_2023, roy_probing_2015}.  However, spin-exchange originates from the Pauli exclusion principle and SE collisions are thus short-range interactions, with a cross-section $\sigma \propto a_0^2$ where $a_0$ is the Bohr radius. As a consequence, they are responsible for relaxations and decoherence, but the resulting interaction is so short-lived that the spins dynamics remains essentially that of independent particles.  Namely, spin noise spectra can be inferred either from one-body dynamics or from mean-field descriptions of the evolution of single-particle operators \cite{mouloudakis_spin-exchange_2021, mouloudakis_interspecies_2023}. Consequently, collisions-induced entanglement was studied theoretically \cite{mouloudakis_spin-exchange_2021} but lacks experimental evidence. Moreover, higher-order spin correlators in non-interacting atomic samples are quite challenging \cite{li_higher-order_2013} and have not yet been observed without the use of external driving fields \cite{li_higher-order_2016}.

In this letter, we report the measurement of a strongly-interacting SNS regime, where previously unobserved spectral features appear in a high-density Rubidium vapor when probing SNS close to an atomic resonance. We analyze both strong detuning and density-dependent changes in the lineshape, that we attribute to long-range dipole-dipole interactions (DDI) between atoms. To support this claim, we perform numerical simulations including DDI in a two-body system to show that this interaction is responsible for the observed changes in the two-atom spin noise spectrum.

\begin{figure}
    \centering
    \includegraphics[width=\columnwidth]{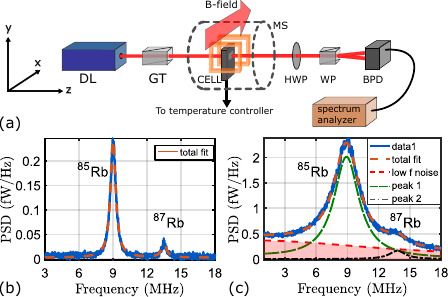}
    \caption{(a) Experimental setup used for SNS experiments. DL : diode laser, GT : Glan-Thompson polarizer, MS : magnetic shielding, HWP : half-wave plate, WP : Wollaston prism, BPD : balanced photodetection. (b,c) Spin noise spectra obtained for a cell temperature (b) $T=90^{\circ}$C and (c)  $T=175^{\circ}$C. The shaded area in (c) shows the low-frequency noise discussed in the text.}
    \label{figure1}
\end{figure}

\paragraph{Experimental results.} We perform spin noise spectroscopy by measuring the stochastic Faraday rotation noise near the $\mathrm{D}_2$ line of Rubidium. The experimental setup is depicted in Fig.\,\ref{figure1} (a). We use an external cavity diode laser to provide a few-mW probe beam at 780\,nm, propagating in the $z$ direction. It is linearly polarized and focused with a diameter $\Phi = 0.3 \,\mathrm{mm}$ into a 1-mm-thick glass cell containing both $^{85}\mathrm{Rb}$ and $^{87}\mathrm{Rb}$ isotopes with natural abundance, and no buffer gas. A pair of coils creates a transverse magnetic field aligned along the $x$ axis, modulating the spin noise resonances at the Larmor precession frequency, which can be tuned in the MHz range. The temperature of the cell is monitored and stabilized using a home-made temperature controller, reaching temperatures as high as $180^{\circ}\mathrm{C}$. The vapor density is then inferred using the Killian's formula \cite{siddons_absolute_2008}. The probe polarization orientation noise is measured with a balanced detection, the output of which is fed in an electronic spectrum analyzer, giving access to the spin noise power spectral density (PSD). 

A typical example of SN spectrum obtained at relatively low density  $N \simeq 10^{12}\,\mathrm{at.cm}^{-3}$ (T = 90$^{\circ}$C) can be seen in Fig.\,\ref{figure1} (b). The probe power is 2\,mW and the magnetic field is $B\simeq 19\,\mathrm{G}$. This spectrum is obtained by tuning the probe laser between the $F=2\rightarrow F'$ and $F=3\rightarrow F'$ hyperfine transitions between the  $5^2S_{1/2}$ ground state and $5^2P_{3/2}$ excited state of $^{85}\mathrm{Rb}$. Defining the detuning as $\Delta = \omega_p - \omega_0$ with $\omega_p$ the probe laser frequency and $\omega_0$ the $F=3\rightarrow F'$ hyperfine transition frequency, this spectrum was obtained for $\Delta/2\pi = 1.5\,\mathrm{GHz}$. This detuning is much larger than the Doppler broadening of the ground-state hyperfine transitions and thus the absorption is negligible. The laser shot noise floor has been subtracted from the data. The spin noise spectrum then shows one SN resonance for each isotope, the width of which is mainly determined by the transit rate of the atoms through the laser beam in ballistic regime: $\gamma_t/2\pi \simeq \bar{v}/(2\pi\Phi) = 250\,\mathrm{kHz}$. Here $\bar{v} = \sqrt{3k_b T/m}$ is the average thermal velocity in the vapor. This simple spectrum is very well understood and arises from the stochastic precession of a single spin in a transverse magnetic field, under relaxation processes such as transit decoherence or SE collisions.

In order to probe higher density regimes, we perform SNS in a 1-mm-thick cell, thin enough to obtain low optical depth even close to resonance and thus reduce light absorption. Consequently, we are able to work at large atomic densities (up to around $4\times 10^{14} \,\mathrm{at.cm}^{-3}$ at $T=180^{\circ}\mathrm{C}$) while keeping the detuning as small as $\Delta/2\pi = 1.5\,\mathrm{GHz}$. This would not be possible using typical cm-long spectroscopy cells. At such densities in our case, absorption is not negligible but the output transmission (typically 25 \%) is still substantial. In this configuration, we report drastic changes in the spin noise lineshapes when the density of the vapor increases. A SN spectrum recorded in similar conditions to that of Fig.\,\ref{figure1}(b) except for a higher temperature of T=$ 175^{\circ}\mathrm{C}$ is shown in Fig.\,\ref{figure1}(c). One can clearly see a large broadening of both spin resonance lines, up to widths of the order of $1\,\mathrm{MHz}$. Additionally, a broad, unexpected low-frequency noise component appears. We quantify the broadening of the peaks as well as this low-frequency noise by using a fit function consisting of the sum of three lines, with lineshapes obtained from the spectral analysis of the stochastic Bloch equation, that can be found in \cite{delpy_compagnon}. These three lines appear in red dashed line (low-frequency noise), green dashed line ($^{85}\mathrm{Rb}$ peak) and black dash-dotted line ($^{87}\mathrm{Rb}$ peak) in Fig.\,\ref{figure1} (c).

\begin{figure}
    \centering
    \includegraphics[width=\columnwidth]{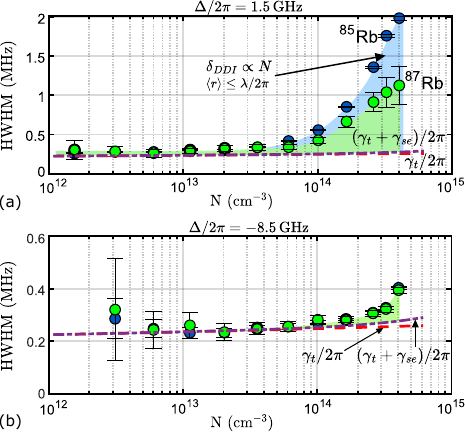}
    \caption{(a) Evolution of the half-width at half maximum (HWHM) of the SN peaks of both isotopes as a function of the Rb vapor density for a probe laser of 2\,mW and a detuning $\Delta/2\pi = 1.5\,\mathrm{GHz}$. The shaded areas mark the part of the linewidth $\delta_{DDI}$ which we attribute to the dipole-dipole coupling. (b) Same measurement for a larger detuning $\Delta/2\pi =-8.5\,\mathrm{GHz}$.}
    \label{figure2}
\end{figure}

More insight into this anomalous lineshape can be obtained by measuring the half width at half maximum (HWHM) of both spin resonance peaks as the temperature of the cell is varied between $80^{\circ}$C and $180^{\circ}$C. This corresponds to an atomic density spanning  more than two orders of magnitude, from $N \simeq 10^{12} \,\mathrm{at.cm}^{-3}$ to $4 \times 10^{14} \,\mathrm{at.cm}^{-3}$. Figure \ref{figure2}(a) shows the resulting trend for both isotopes using the same $\Delta/2\pi = 1.5\,\mathrm{GHz}$ optical detuning. The dashed red line is the value of $\gamma_t/2\pi$, which corresponds to a width limited by the transit of the atoms into the laser beam. The purple dash-dotted line is the sum of the transit and the SE collisional rates $(\gamma_t+\gamma_{se})/2\pi$, using the theoretical expression $\gamma_{se} = K N_{at}\sigma_{se}\bar{v}/2\pi$, where $\sigma_{se} = 2\times 10^{-14}\,\mathrm{cm}^2$ is the spin-exchange cross section and $K\simeq 0.4$ a factor arising from the nuclear angular momentum of Rubidium \cite{katsoprinakis_measurement_2007}. We recall that these two relaxation processes determine the one-body dynamics in typical SNS experiments. In our case however, the observed HWHM increases from around $\gamma_t/2\pi = 250\,\mathrm{kHz}$ at low density up to almost 2 MHz at the highest temperature for the $^{85}\mathrm{Rb}$ noise peak. Strikingly, none of the experimental data can be explained by the relaxation processes discussed above as soon as the vapor density approaches $N = 10^{14}\,\mathrm{at.cm}^{-3}$. The remainder of the linewidth, denoted $\delta_{DDI}$ and highlighted in shaded areas in Fig.\,\ref{figure2}(a), cannot either be explained by power broadening induced by the probe light.
Indeed, the ground-state population excitation rate is of the order of $\Gamma_p \propto \Gamma \dfrac{\Omega^2}{\Omega^2+\Delta^2}$, with $\Gamma$ the optical dipole relaxation rate and $\Omega$ the Rabi frequency. This amounts to $\Gamma_p/2\pi \simeq 30\,\mathrm{kHz}$ in our case, far from the measured value of $\delta_{DDI}$. 

Consequently, we attribute this unusually large linewidth to the resonant dipole-dipole interaction (DDI) induced by the probe light between the atoms in the vapor. To support this idea, the same measurements were performed at a larger (absolute) detuning of $\Delta/2\pi = -8.5\,\mathrm{GHz}$,  as shown in Fig.\,\ref{figure2}(b). They reveal a much smaller broadening, due to the atomic dipole being more weakly driven at such a large detuning. However, although the acquired signal is much weaker at this detuning, leading to large fit uncertainties at lower density, the additional broadening $\delta_{DDI}$ is still measurable for the same range of temperatures. The critical density of $N= 10^{14}\,\mathrm{at.cm}^{-3}$ highlighted earlier can then be interpreted as corresponding to an average interatomic distance of $\langle r \rangle  \simeq 120\,\mathrm{nm} \simeq \lambda/2\pi$ at which atoms start to be radiatively coupled \cite{lehmberg_radiation_1970}. To further explain how such lineshapes arise from dipole-dipole coupling, we now give a microscopic description of the system and perform numerical simulations.

\begin{figure}[t]
    \centering
    \includegraphics[width=0.95\columnwidth]{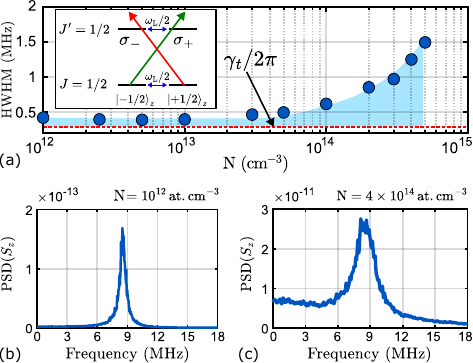}
    \caption{(a) Simulated evolution of the HWHM of the SN peak as a function of atomic density. Parameters : $\Omega/2\pi = 150\,\mathrm{MHz}\,(P = 2\,\mathrm{mW})$, $\Delta/2\pi = 300\,\mathrm{MHz}$. (b,c) Examples of spin noise spectra simulated for densities of (b) $10^{12}\, \mathrm{at.cm}^{-3}\,(T\simeq 75^{\circ}\mathrm{C})$ and (c) $4\times 10^{14}\, \mathrm{at.cm}^{-3}\,(T\simeq 180^{\circ}\mathrm{C})$.}
    \label{figure3}
    \vspace{-0.2cm}
\end{figure}
\paragraph{Numerical simulations.} We now provide a simple model to study the effect of the DDI on the dynamics of the ground-state coherences relevant to SNS. We consider the simplest possible toy model  that still reproduces the experimental observations. First, the fact that $\gamma_t \gg \gamma_{se}$ leads to neglect inter-hyperfine coherences arising from spin-exchange and thus to overlook the hyperfine structure of the atoms. Second, as we are interested only in the spin orientation noise, we limit ourselves to the case of a simple $J= 1/2$ ground and excited states. The corresponding energy structure is depicted in the inset of Fig.\,\ref{figure3}(a), with a quantization axis chosen to be the light propagation axis $z$. With this choice, Zeeman sublevels are coupled by the transverse magnetic field with a strength $\hbar \omega_L/2 $, being the Larmor frequency of the ground-state coherences. The $\sigma_+$ (resp. $\sigma_-$) component of the light couples the ground $\ket{g,-1/2}_z$ Zeeman sublevel to the excited $\ket{e,+1/2}_z$ one (resp. $\ket{g,+1/2}_z$ to $\ket{e,-1/2}_z$). These two couplings determine the single-particle Hamiltonian $H_0$. 

Now, considering two of these atoms, the evolution of the two-body density matrix including the dipole-dipole interaction is obtained by solving the Master Equation (ME) 
\begin{equation}
     \dot{\rho} = -\dfrac{i}{\hbar} \left[H, \rho\right] + D[\rho] +f\ ,
     \label{eq1}
 \end{equation} where $f$ is a noise operator similar to the one discussed in \cite{liu_spin-noise_2023, liu_birefringence_2022}.
Here, $H = \sum_{i=1,2} H_0^{(i)} + V_{dd}$ is the total Hamiltonian acting on the two-body Hilbert state, and $D[\rho]$ the two-body decoherence operator. This master equation can be carefully derived by considering a system of two atoms interacting with the vacuum modes of the electromagnetic field \cite{lehmberg_radiation_1970, agarwal_quantum_2012, le_kien_nanofiber-mediated_2017, reitz_cooperative_2022}. Considering the latter as a bath, tracing out its degrees of freedom, one eventually obtains the above equation of motion. This yields for the interaction Hamiltonian $V_{dd} = \hbar \sum_{\substack{i\neq j\\\lambda,\lambda'}}\zeta_{\lambda,\lambda'}(r_{ij}) \frac{D_+^{\lambda,i} D_-^{\lambda',j}}{d_0^2}$. Here, $D_{+}^{\lambda,i}$ (resp. $D_{-}^{\lambda,i}$) is the excitation (resp. de-excitation) part of the dipole operator of atom $i$ and of polarization component $\lambda =\sigma_+, \sigma_-,z$. We denote $d_0$ the dipole element for the $J\rightarrow J'$ transition and $\Gamma_0$ the excited states population relaxation rate. The quantity $\displaystyle{\zeta_{\lambda,\lambda'}(r_{ij}) = -\dfrac{3 \Gamma_0}{4} \, \left(\bs{\varepsilon}_{\lambda}\right)^*\cdot\mathrm{Re}\qty(\rttensor{G}(\mathbf{r}_i, \mathbf{r}_j, \omega_0))\cdot\bs{\varepsilon}_{\lambda'}}$, where $\displaystyle{\rttensor{G}(\mathbf{r}_i, \mathbf{r}_j, \omega_0) = \qty(\rttensor{\identity}+\dfrac{\bs{\nabla}\bs{\nabla}}{k^2} ) \dfrac{e^{i k_0 r_{ij}}}{k_0 r_{ij}}}$ is the electromagnetic Green's dyadic, gives the strength of the radiative coupling between a dipole in $\mathbf{r}_i$  polarized along $\bs{\varepsilon}_{\lambda'}$ and a second one placed in $\mathbf{r}_j$ oscillating in the direction of $\bs{\varepsilon}_{\lambda}$. Here we denoted $k_0 = \omega_0/c$ the light wave vector and $r_{ij} = \vert \mathbf{r}_{i}- \mathbf{r}_{j}\vert$ the interatomic distance. It is noteworthy that this coupling depends strongly on $r_{ij}$, as $\rttensor G((\mathbf{r}_i, \mathbf{r}_j, \omega_0) \propto 1/r_{ij}^3$  in the near field \cite{browaeys_experimental_2016, agarwal_quantum_2012, varada_two-photon_1992, lehmberg_radiation_1970, reitz_cooperative_2022}.  The spherical angles $\theta, \phi$ between the two atoms are also encapsulated in the Green's tensor. The decoherence matrix takes into account both the transit relaxations and the atomic optical decoherence due to the coupling with the electromagnetic bath \cite{agarwal_quantum_2012}.

Numerically solving the ME, one can obtain a set of simulated time traces for the two-atom density matrix $\rho$. To account for the thermal motion of the atoms in the vapor, we vary periodically the geometry of the two-body system by randomly picking new spherical coordinates after a correlation time $\tau_c$. The interatomic distance follows the nearest neighbor distribution in a dilute vapor explicited in \cite{ chandrasekhar_stochastic_1943, falvo_double-quantum_2023, yu_long_2019}. Its mean value is given by the chosen atomic density. The angular coordinates follow a uniform probability. For each time trace, the ensemble average $\langle S_z\rangle = \mathrm{Tr}(\rho S_z)$ of the spin operator $S_z=\sum_{i=1,2}s_z^{(i)}$ of the two-body system  is then computed at each time, along with its Fourier Transform $\hat{F}[S_z] (\omega)$. One then accesses the two-body spin noise $\mathrm{PSD} \equiv \langle \vert \hat{F}(\omega)^2\vert \rangle$, $\langle ... \rangle $ denoting the average over all simulated samples. Figures \ref{figure3}(b) and (c) show examples of SN spectra simulated for $N = 10^{12}\, \mathrm{at.cm}^{-3}$ and $10^{14}\, \mathrm{at.cm}^{-3}$ respectively, with a Rabi frequency $\Omega/2\pi = 150\,\mathrm{MHz}$ (matching the 2\,mW probe power used experimentally), a detuning $\Delta/2\pi = 300\,\mathrm{MHz}$ and a Larmor frequency $\omega_L/2\pi = 9\,\mathrm{MHz}$. The 300\,MHz detuning, far from the 1.5\,GHz experimental one, was chosen to keep the ratio $\Delta/\Gamma$ constant: due to the large Doppler inhomogeneous linewidth, this is the simplest way to take into account a saturation phenomenon with the good order of magnitude.

The shapes of both spectra match very well the experimental results of Figs.\,\ref{figure1} (b) and (c). At low density, the width of the peak equals the transit decoherence rate, and corresponds to the usual single-body dynamics. At high density the simulations precisely reproduce the broadening of the peak up to the MHz range, as well as the appearance of a broad background noise centered near zero frequency. Figure \ref{figure3}(a) shows the evolution of the HWHM of the simulated SN peak with the density. As the latter increases, the average interatomic distance decreases and the resonance broadens. These numerical results are close to the experimental ones highlighted in Fig.\,\ref{figure2}(a). This agreement shows that a two-body description of the spin ensemble successfully explains lineshapes that are unexpected in a non-interacting spin ensemble.

\begin{figure}
    \flushleft
    \includegraphics[width=0.95\columnwidth]{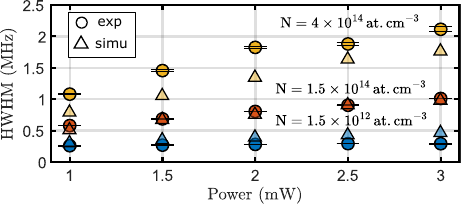}
    \caption{Circles: experimental broadening of the $^{85}\mathrm{Rb}$ spin noise resonance as a function of laser power, for three different atomic densities. Triangles: Corresponding numerical simulations.}
    \label{figure4}
\end{figure}

\paragraph{Discussion.} We now discuss how an exchange of atomic excitations can actually induce changes in the dynamics of the ground-state coherences and populations. First, the action of the near-resonant laser is to induce coherences between ground and excited states \cite{chalupczak_near-resonance_2011, horn_spin-noise_2011, delpy_spin-noise_2023}, which can then give rise to dipole-dipole coupling between atoms at first order (resonant interaction). Indeed, Fig.\,\ref{figure4} shows the evolution of the HWHM of the $^{85}\mathrm{Rb}$ spin noise as the laser power increases, for temperatures of $T=90^{\circ}$C ($N_{at} = 1.5\times10^{12}\,\mathrm{at.cm}^{-3}$), 160$^{\circ}$C ($N_{at} = 1.5\times10^{14}\,\mathrm{at.cm}^{-3}$), and 180$^{\circ}$C ($N_{at} = 4\times10^{14}\,\mathrm{at.cm}^{-3}$). At low temperatures, the linewidth seems almost independent of the laser power, since the distance between the atoms is too large for the interaction to be relevant. Now, as soon as $\langle r\rangle < \lambda/2\pi$ ($T> 150^{\circ}$C), the interaction strength becomes non negligible and one can clearly see a net increase in the peak width as the laser drives the dipole more efficiently. Such observations are very well reproduced by our model (see the triangles in Fig.\,\ref{figure4}). The discrepancy between experiments and simulations being larger at the highest density may suggest the limit of our two-body model and the need to include three or more particles.

The observed broadening can be explained in the following manner: in presence of the DDI, the eigenenergies of the two-body lower Zeeman manifold differs from the case where the atoms are uncoupled. The changes in the spin precession frequencies according to these new two-body eigenenergies induce a splitting of the SN lines, as shown in \cite{delpy_compagnon}. This splitting is eventually blurred because of the fluctuations in the interatomic distance and orientation due to the atomic thermal motion, to yield a single, broadened peak. In this regard, the broadening $\delta_{DDI}$ should be viewed as an inhomogenous broadening of the spin noise peaks and not as an additional decoherence rate. Furthermore, the finite lifetime of the bipartite system (represented by the geometry correlation time in our model) also broadens the spin noise lineshape. However, the strong dependence of the linewidth on the laser power suggests that the inhomogeneous broadening actually dominates the one induced by this finite lifetime.

\begin{figure}[t]
    \flushleft
    \includegraphics[width=\columnwidth]{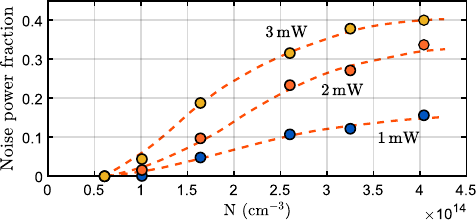}
    \caption{Evolution of the fraction of the total noise power contained in the low-frequency component as a function of the vapor density, for three different probe powers. Dashed lines are guides to the eye. }
    \label{figure5}
\end{figure}

Finally, analyzing the broad, low-frequency noise component appearing at higher densities is also a fruitful approach to characterize the correlated dynamics of the two-body system. Figure \ref{figure5} shows that the ratio of the noise power contained in this component (integrated from our fitting procedure) over the total spin noise power increases quickly with the vapor density as soon as the DDI appears, and that this increase is faster for larger laser powers. This shows that this extra noise is another signature of the light-induced dipole-dipole coupling. Taking advantage of the time traces obtained from our simulations (see \cite{delpy_compagnon}), we suggest that this extra noise actually originates from the spin fluctuations due to the changes in the spatial arrangement of the binary. For a given conformation, the two atoms evolve towards a steady-state characterized by an intrinsic value of the two-particles operator $\braket{S_z}^{(st)}$, before the moving atoms eventually separate after a finite time. The low-frequency component thus witnesses the spin noise created by the DDI being a noisy process on its own. Its lineshape must be the spectral signature of the time dynamics of the binary. The fit of the data shown in Fig.\,\ref{figure1}(c) gives a HWHM for this line of $\gamma/2\pi = 14.1\pm0.8\,\mathrm{MHz}$. Following the Wiener-Khintchine theorem, this gives a correlation time of the two-body evolution of $\tau_c = 11.2\pm 0.6\,\mathrm{ns}$. 
This corresponds to an effective dipole-dipole interaction range of $a_c = \bar{v}\tau_c = 4.1\pm0.2$ \textmu m, consistent with the recent measurement of a DDI interaction range $a_c = 6.1$ \textmu m in Rb using two-dimensional spectroscopy \cite{liang_optical_2021, yu_long_2019, gao_probing_2016}.
As a comparison, this interaction range is two orders of magnitudes larger than the one appearing in the self-broadening of a hot vapor \cite{lewis_collisional_1980, silans_determination_2018, weller_absolute_2011}. Indeed, in that case, the relevant impact parameter would be the Weisskopf radius  $\rho_w \simeq 20\,\mathrm{nm}$ at 300K , for which a DDI collision dephases two dipoles by around $1\,\mathrm{rad}$ \cite{Akulin_Karlov_1992, leegwater_self-broadening_1994, ali_theory_1965,sautenkov_dipole-dipole_1996}.

To summarize, we have demonstrated a drastic influence of long-range optical interactions on the spin dynamics of a hot atomic vapor. We have isolated a high-density, optically driven regime of SNS in which the spin fluctuations can no longer be attributed to single-particle dynamics. Using a microscopic model of two atoms coupled by resonant dipole-dipole interaction, we have successfully reproduced the observed lineshapes, proving that we are indeed observing a genuine two-body spin noise. Finally, we have characterized the lifetime of the correlated two-atom system, showing an interaction range of the order of a few \textmu m. This interaction range $a_c$ being greater than the average interatomic distance at high densities, the two-body approximation must be questioned and future prospects include the study of a three or more body dynamics. Such collective effects in spin noise also opens the way to characterize optical entanglement using SNS, as well as the study of higher order spin noise spectroscopy in strongly interacting atomic vapors.

\acknowledgments
The authors are happy to thank Antoine Browaeys, Thomas F. Cutler, Ifan G. Hughes, and Charles S. Adams, for their help with thin cells. We also benefited from the help of Yassir Amazouz and Garvit Bansal for their early interest in the project,  from S\'ebastien Rousselot for technical assistance, and Thierry Jolic\oe ur for some help with simulations. The authors acknowledge funding by the Labex PALM.

\selectlanguage{english}

\bibliographystyle{apsrev4-1}

\appendix

\end{document}